\documentclass[useAMS,usenatbib]{mn2e}
\usepackage{graphicx}
\usepackage{float}
\usepackage{mathtools}
\usepackage{color}
\usepackage{tabularx}
\usepackage{sidecap}
\usepackage{longtable}
\def\kms{km\,s$^{-1}$}

\def\Ha{H$\alpha$}
\def\Hb{H$\beta$}

\def\CaII{Ca\,{\sc ii}}
\def\SrII{Sr\,{\sc ii}}

\def\OI{O\,{\sc i}}

\def\OIII{O\,{\sc iii}}

\def\SiII{Si\,{\sc ii}}
\def\SiI{Si\,{\sc i}}
\def\SrII{Sr\,{\sc ii}}
\def\ScII{Sc\,{\sc ii}}
\def\TiII{Ti\,{\sc ii}}

\def\ScII{Sc\,{\sc ii}}
\def\HeI{He\,{\sc i}}

\def\OIII{O\,{\sc iii}}
\def\FeII{Fe\,{\sc ii}}

\def\MgI{Mg\,{\sc i}}

\def\NaI{Na\,{\sc i}}

\def\msun{M$_{\odot}$}
\def\lesssim{\mathrel{\hbox{\rlap{\hbox{\lower4pt\hbox{$\sim$}}}\hbox{$<$}}}}
\def\gtrsim{\mathrel{\hbox{\rlap{\hbox{\lower4pt\hbox{$\sim$}}}\hbox{$>$}}}}
\onecolumn
\title[SN~2013by]
{Supernova 2013by: A Type IIL Supernova with a IIP-like light curve drop\thanks{
This paper is based on observations gathered 
with: the LCOGT network of telescopes, the 6.5 meter Magellan Telescopes,  the Swope 1 meter telescope, and the $Swift$ telescope.
}}
\author[Valenti et al.]{S. Valenti$^{1,2}$\thanks{e--mail: svalenti@lcogt.net}, D. Sand$^{3}$, M. Stritzinger$^{4}$, 
D.~A. Howell$^{1,2}$, I. Arcavi$^{1,5}$, C. McCully$^{1,2}$,
\and 
M. J. Childress$^{6,7}$,  E.Y. Hsiao$^{4,8}$, C. Contreras$^{4,8}$, N. Morrell$^{8}$, M. M. Phillips$^{8}$,  
\and
M. Gromadzki$^{9,10}$, R. P. Kirshner$^{11}$,  G. H. Marion$^{11,12}$ \\
$^{1 ~}$ Las Cumbres Observatory Global Telescope Network, 6740 Cortona Dr., Suite 102, Goleta, CA 93117, USA \\
$^{2 ~}$ Department of Physics, University of California, Santa Barbara, Broida Hall, Mail Code 9530, Santa Barbara, CA 93106-9530, USA \\
$^{3 ~}$ Physics Department, Texas Tech University, Lubbock, TX , 79409, USA\\ 
$^{4~}$ Department of Physics and Astronomy, Aarhus University, Ny Munkegade 120, DK-8000 Aarhus C, Denmark \\
$^{5~}$ Kavli Institute for Theoretical Physics, University of California, Santa Barbara, CA 93106, USA\\
$^{6~}$ Research School of Astronomy and Astrophysics, Australian National University, Canberra, ACT 2611, Australia \\
$^{7~}$ ARC Centre of Excellence for All-sky Astrophysics (CAASTRO); Australian National University; Canberra, ACT 2611,
 Australia \\
 $^{8~}$  Carnegie Observatories, Las Campanas Observatory, Colina El Pino, Casilla 601, Chile \\
 $^{9~}$ Millennium Institute of Astrophysics, Sotero Sanz 100, Oficina 104, Providencia, Santiago \\
$^{10}$ Instituto de F{\'i}sica y Astronom{\'i}a, Universidad de Valpara{\'i}so, Av. Gran Breta\~{n}a 1111, Playa Ancha, Casilla, 5030, Chile \\
$^{11}$ Harvard-Smithsonian Center for Astrophysics, 60 Garden Street, Cambridge, MA 02138, USA )\\
$^{12}$ University of Texas at Austin, 1 University Station C1400, Austin, TX, 78712-0259, USA \\
}
\begin{document}

\date{Accepted .....; Received ....; in original form ....}


\maketitle

\begin{abstract}
We present multi-band ultraviolet and optical light curves, as well as
visual-wavelength and near-infrared spectroscopy of the Type II linear
(IIL) supernova (SN) 2013by. We show that SN~2013by and other SNe IIL
in the literature, after their linear decline phase that start after 
maximum, have a sharp light curve decline similar to that seen 
in Type II plateau (IIP) supernovae.  This light curve feature has rarely been
observed in other SNe~IIL due to their relative rarity and the
intrinsic faintness of this particular phase of the light curve.  
We suggest that the presence of this drop could be used as a 
physical parameter to distinguish between subclasses of SNe~II, 
rather than their light curve decline rate shortly after peak.  
Close inspection of the spectra of SN~2013by indicate asymmetric 
line profiles and signatures of high-velocity hydrogen.  
Late ($\sim$ 90 days after explosion) near-infrared spectra of 
SN~2013by exhibit oxygen lines, indicating significant mixing within 
the ejecta. From the late-time light curve,
we estimate that 0.029 \msun{} of $^{56}$Ni was synthesized during the
explosion.  It is also shown that the $V$-band light curve slope is
responsible for part of the scatter in the luminosity ($V$ magnitude
50 days after explosion) vs. $^{56}$Ni relation.  Our observations of
SN 2013by and other SNe IIL through the onset of the nebular phase
indicate that their progenitors are similar to those of SNe IIP.
\end{abstract}

\begin{keywords}
supernovae: general -- supernovae: SN~2013by,  --  galaxies:  
\end{keywords}

\section{Introduction}
\label{parintroduction}
Type~II supernovae (SNe~II) have historically been divided into the
Type~IIL (linear) and Type~IIP (plateau) subclasses based on the shape
of their light curves in the weeks following explosion\footnote{Note
  that we are not considering Type~IIn nor Type~IIb in this work, even
  though there are still several open issues on where these objects
  are situated in terms of SN~II diversity.}  \citep{Barbon1979}.
SNe~IIP light curves have a clear plateau phase, where the SN
brightness stays nearly constant (in the optical) for roughly 100
days.  The plateau phase is mainly powered by a moving hydrogen
recombination front that travels through the hydrogen-rich material
that is ejected and ionized during the explosion
\citep{Woosley1987}. A prototypical example of a Type IIP SN is
SN~1999em \citep{Hamuy2003}.  SNe~IIL exhibit a linear decay that
starts soon after peak brightness, They are more rare than SNe IIP
and, because light curves of Type IIL SNe decline faster than light
curves of Type IIP SNe, only a handful of SNe IIL have been followed
for more that three months after peak brightness.  Prototypical
example of Type IIL SNe are SN~1979C and SN~1980K \citep[see
][]{Filippenko1997}.

SNe~IIL are on average more luminous than SNe~IIP by $\sim$1.5 mag
\citep{Patat1993, Patat1994, Anderson2014, Sanders2014, Faran2014a}.
Spectroscopically, SNe~IIL have on average redder continua
and have higher oxygen to hydrogen ratio as compared to ordinary SNe~IIP
\citep{Faran2014a}.  SNe~IIL also exhibit higher expansion velocities
at early times \citep{Faran2014a}, and less pronounced P-Cygni \Ha{}
profiles \citep{Gutierrez2014}.

Given these differences, there is a general consensus that the
progenitor stars of SNe~IIL have a relatively small amount of hydrogen
in their envelopes, while SNe~IIP more likely originate from
hydrogen-rich stars. It is not currently known whether the progenitors
of SNe~IIL gradually lose their hydrogen layer, creating a continuous
class extending from SNe~IIL to SN~IIP, or if there is a specific
mechanism that creates distinct classifications as has been previously
suggested \citep{Barbon1979,Patat1993}.

This traditional classification scheme has been supported in recent 
compilations of  SN~II light curves by \cite{Arcavi2012} and 
\cite{Faran2014a}. However, \cite{Anderson2014} and 
\cite{Sanders2014} have shown that the historical distinction  
between the two classes could be due to the small number of well observed
SNe~IIL.  They have also suggested that 
the historical distinction between IIP and IIL, based on the presence of a plateau 
 or a linear decay in the light curve, is insufficient for a complete mapping 
 of SN~II diversity. \cite{Anderson2014} suggested that if all SNe~IIL are followed 
for a long enough time, they will exhibit, after the linear decay, a significant drop
in their light curves. This would provide further evidence that IIP's and IIL's 
share the same underlying physics. However, the majority of  SNe~IIL 
recently presented by \cite{Anderson2014} and \cite{Sanders2014} 
have only been followed for a limited amount of time ($\lesssim$ 
70-80 days from discovery). 

Here we present detailed ultraviolet and optical broad-band photometry 
of the Type~IIL SN~2013by, which covers the flux evolution  of this 
object for over 150 days. Also presented are four visual-wavelength 
spectra and three near-IR  (NIR) spectra. The data presented have been 
obtained  by the Las Cumbres Observatory Global Telescope (LCOGT) 
network \citep{Brown2013a}, the   {\em Carnegie Supernova Project} 
\citep[CSP; ][]{Hamuy2006}, and with the UVOT camera aboard the 
\emph{Swift} X-ray telescope  \citep{Burrows2005,Roming2005}.

The organisation of this paper is as follows. In Section~\ref{sec:13bydata} 
we present the spectroscopic and photometric observations, and 
briefly characterise the data reduction process. Presented in 
Section~\ref{sec:lightcurve} are the photometric data of 
SN~2013by, while in Section~\ref{sec:spectra} we analyse the 
optical and NIR spectra of SN~2013by.
In Section \ref{sec:nickel} we discuss the amount of $^{56}$Ni 
produced in SN~2013by, and compare our results with the 
previous works of \cite{Hamuy2003} and \cite{Spiro2014}.
Our results are summarized in Section \ref{sec:discussion}.

\section{Observations}
\label{sec:13bydata}

SN~2013by was discovered by the Backyard Observatory Supernova Search
(BOSS) on 2013 April 23.542 (UT), with coordinates at
$\alpha$=16h59m02s.43, $\delta$=$-$60d11'41".8 \citep{Parker2013}.
The SN is located 3$\arcsec$ West and 76$\arcsec$ North of the nucleus
of the galaxy ESO 138$-$G10. The the NASA Extragalactic Database
(NED)\footnote{http://ned.ipac.caltech.edu/} distance, corrected for
Local-group infall towards the Virgo cluster and assuming $H_0 =
73\pm5$ \kms\,Mpc$^{-1}$, is D=14.8$\pm$1 Mpc (distance modulus =
30.84 $\pm$0.15).  This value is adopted throughout this work.  The
closest available pre-discovery limit is not stringent \citep[on April
  1.554 UT, ][]{Parker2013}, but early spectroscopic and photometric
follow-up are consistent with a supernova discovered a few days or
less after explosion. In what follows we adopt April 21.5 (UT), i.e.,
JD = 2456404$\pm$2 days, as the explosion epoch.  Spectroscopic
confirmation as a young SN~II came from both optical and NIR
observations \citep{Parker2013}, along with a tentative classification
as a Type IIL/IIn.  Based on the light curve, SN~2013by, in the
traditional schema, is a typical SN~IIL (see Section
\ref{sec:lightcurve}). However X-ray emission has also been detected
for SN~2013by with $Swift$ \citep{Margutti2013}, supporting the idea
that SN~2013by may have experienced moderate interaction with
circumstellar material (CSM) during early phases (see
Section~\ref{sec:spectra}).

Photometric monitoring in $BVgri$ of SN~2013by with the LCOGT 1 m
telescope network began on 2013 April 24 (UT), and continued every 2-3
nights (52 epochs of data were collected) for more than 150 days, well
after the light curve settled onto the $^{56}$Co decay tail.  The
LCOGT science images were reduced using a custom pipeline that
performs point spread function (PSF) fitting and a low order
polynomial fit to remove any background contamination \citep[see
][]{Valenti2014}. SN~2013by exploded close (but not coincident) to two
point-like sources that are $\geq$ 2 magnitudes fainter than our last
detection.
 
Additional imaging was obtained from {\it Swift} and the CSP. The {\it
  Swift} data (15 epochs) was reduced following the standard
procedures described by \cite{Brown2009a}. While the PSF fitting
technique gives a good result for ground-based data of SN~2013by, {\it
  Swift} magnitudes are computed using aperture photometry. By
comparing initial {\it Swift} results with ground based measurements,
host galaxy contamination was evident in the {\it Swift} data.  This
motivated a {\it Swift} Target of Opportunity program (PI S.Valenti)
to re-image the field of SN~2013by one year after discovery in order
to properly remove background host-galaxy contamination as prescribed
by \cite{Brown2009a}.
 
The CSP obtained 17 epochs of science images using the SITe3 CCD
camera along with a set of $ugriBV$ filters attached to the Swope 1 m
telescope located at Las Campanas Observatory (LCO). These images were
reduced using the methodology described in \cite{Contreras2010} and
\cite{Stritzinger2011}. The CSP, LCOGT and {\it Swift} photometry are
in good agreement except the {\it Swift} $V$ band that is
systematically 0.1 magnitude fainter that the LCOGT and CSP
photometry.  Given the good photometric coverage in $V$ band with CSP
and LCOGT data, we did not investigate this systematic difference
further.

Four epochs of visual-wavelength spectra of SN~2013by were obtained
with WifeS \citep{Dopita2007} on the ANU 2.3m Telescope, and three
epochs of NIR spectra with FIRE \citep{Simcoe2013} on the Magellan 6.5
m Baade telescope (see Table \ref{tab1}).  The optical spectra were
reduced with PyWiFes as described by \cite{Childress2013}, while the
FIRE spectra were reduced using the IDL pipeline \texttt{Firehose}.
The \texttt{Firehose} pipeline performed the following steps: flat
fielding, wavelength calibration, sky subtraction, spectral tracing
and extraction and flux calibration.

\section{Light curves}
\label{sec:lightcurve}
\begin{figure*}
\begin{center}
  \includegraphics[width=16cm,height=8cm]{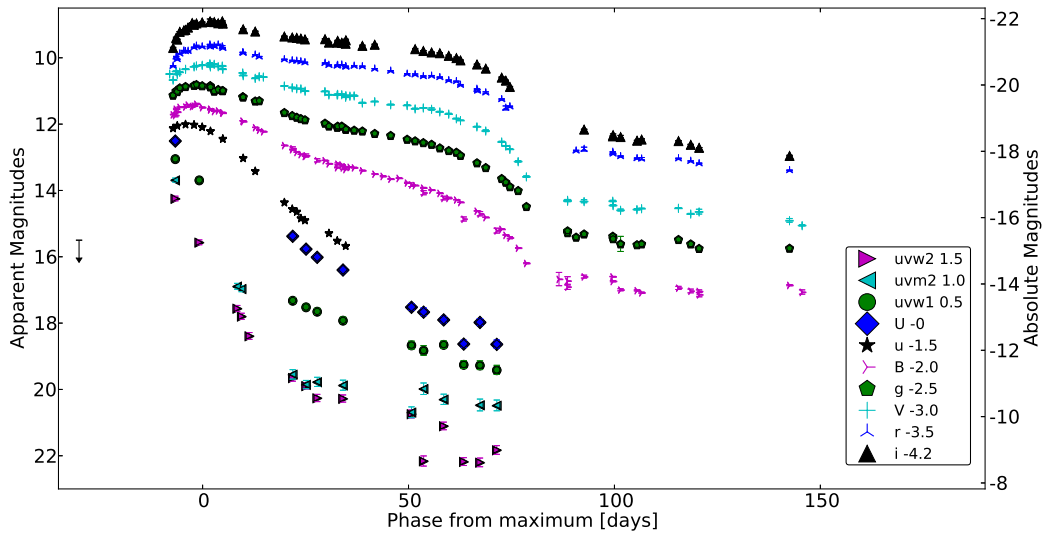}
  \caption{Ultraviolet and optical light curves of SN 2013by from 
  observations obtained by  LCOGT, CSP  and {\em Swift}.  
  From maximum light to $\sim$ 65 days after maximum, SN~2013by 
  experiences a nearly linear decline, with a decline rate per 50 days  of $s50_V=1.46$ mag. 
  The $V$-band light curve then abruptly drops until a new, slow decline is established 
  at $\sim$ +90 days from maximum, presumably associated with the radioactive decay tail of $^{56}$Co $\rightarrow$ $^{56}$Fe.}
  \label{fig:lc}
\end{center}
\end{figure*}

The multi-band  light curves of SN~2013by are shown in Figure~\ref{fig:lc}, 
while the corresponding photometric data are tabulated in Table~\ref{tab2}. 
After inspecting the light curves, we see familiar features including a 
short rise ($\sim$ 10 days) to maximum, followed by a linear phase 
lasting 65 days (from maximum). 
After the linear phase, the light curve of SN 2013by shows a clear drop 
until it sits on the radioactive decay tail of $^{56}$Co to $^{56}$Fe.

\begin{figure*}
\begin{center}
  \includegraphics[width=16cm,height=8cm]{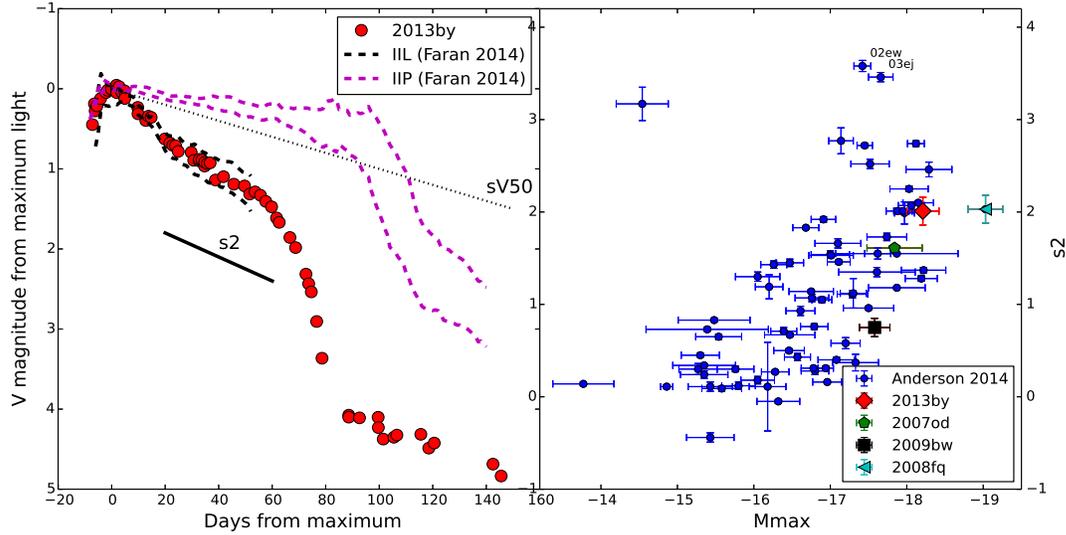}
  \caption{\textit{left panel}: SN~2013by compared with template light curves from 
  \protect\cite{Faran2014b}. The s2 parameter  is the $V$-band 
  magnitudes per 100 days of the second, shallower slope observed in the light curve as defined by \protect\cite{Anderson2014}.  \textit{right panel}: Absolute $V$-band magnitude of SN~2013by vs. s2, compared to objects from \protect\cite{Anderson2014}. SN~2009bw, SN~2007od and SN~2008fq have been also added for comparison. 
  }
  \label{fig:13byIIPIIL}
\end{center}
\end{figure*}

In order to determine which type of SN is SN~2013by, we compare its
$V$-band light curve with templates presented by \cite{Faran2014a}
(see left panel of Figure~\ref{fig:13byIIPIIL}).  SN~2013by lies in
the middle of the SN~IIL templates of \cite{Faran2014a}.  For
completeness we also compare SN~2013by with the CSP sample of SNe~II
published by \cite{Anderson2014} (see right panel of
Figure~\ref{fig:13byIIPIIL}). The \emph{s2} parameter, used by
\cite{Anderson2014} to quantify the slope of the plateau, is the
$V$-band magnitude decline per 100 days measured in the second part of
the plateau (see black line in left panel of
Figure~\ref{fig:13byIIPIIL}).  \cite{Faran2014a} use a different
parameter. Specifically, they use the magnitude decline in 50 days
computed between maximum light and 50 days after explosion
($s50_V$)\footnote{The reader should be aware that since the rise time
  of SNe~II is often not well defined and can takes several days from
  explosion, the $s50_V$ parameter is usually computed within a range
  of $\sim$ 40--45 days.}.  In order to avoid the proliferation of
parameters to characterize SNe~II, we adopt the $s50_V$ in this
paper. \cite{Faran2014a} define all SNe~II with $s50_V>0.5$ mag as
Type~IIL events.  The decline rate for SN~2013by is $s50_V=1.46 \pm
0.06$ mag.

\begin{figure*}
\begin{center}
  \includegraphics[width=16cm,height=8cm]{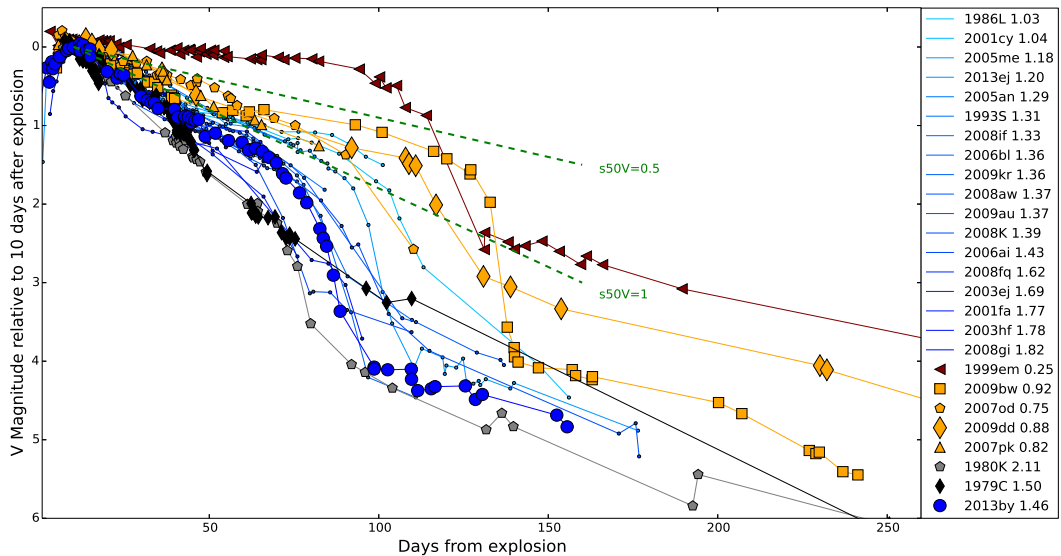}
  \caption{$V$-band light curves of a compilation of SNe~II from the literature, 
  along with SN~2013by.  Each  object is listed and color 
  coded in the right panel along with its measured $s50_{V}$ parameter.  
  We have chosen objects primarily with $s50_V > 1.0$ to highlight the 
  fastest declining, SN IIL-like events.  We have also included several objects 
  with $0.5< s50_V <1.0$, which are plotted in orange and the prototype of 
  Type~IIP SN~1999em.  
  Note that all of  the high $s50_{V}$ objects have a steep drop-off from their linear `plateau' 
  phase at around $\sim$80-120 days, before the radioactive decay tail powers the light curve.  }
  \label{fig:lcV}
\end{center}
\end{figure*}

In this schema SN~2013by should be labeled as a SN~IIL.  However, the
presence of a drop in magnitude at the end of the hydrogen
recombination phase is usually considered the defining feature of a
SN~IIP.  Is the light curve drop-off in SN~2013by atypical for a
SN~IIL?  To answer this question, we have compiled a large sample of
SNe~IIL from the literature, and systematically measured from their
$V$-band light curves the parameter $s50_V$.  Among these, we show a
subset with $s50_V > 1.0$ (typically classified as SN~IIL, as they are
the fastest decliners) in Figure~\ref{fig:lcV}. We also show a handful
of objects with $0.5< s50_V <1.0$ (orange symbols; SNe~2007od, 2007pk,
2009bw, 2009dd). If we follow the definition of \cite{Faran2014a},
where SNe~IIL decline with $s50_V > 0.5$, SN~2007od, SN~2007pk and
SN~2009bw should also be classified as SNe~IIL. 
These SNe~II are as luminous as SNe~IIL \citep[$M_{IIL}=-17.44$~mag][]{Li2011}
\citep[$M^{07od}_{V}=-17.4~$mag and $M^{09bw}_{V}=-17.2$ mag,][]{Inserra2013}.  
For comparison SNe~IIP are fainter ($M_{IIP}=-15.66$ mag), \citep{Li2011}.  
These SNe have been studied in detail by \cite{2011MNRAS.417..261I} and
\cite{2012MNRAS.422.1122I}. They identify high velocity \Ha{}
absorption in the spectra ($\sim$ 13,000-15,000 \kms{}) and suggest
this is evidence for \emph{moderate interaction} with CSM.
\cite{Chugai2007} have shown that the presence of high-velocity
features in Type II SNe can be indeed interpreted as interaction
between rapidly expanding SN ejecta and circumstellar material (CSM).

The SNe in Figure~\ref{fig:lcV} are predominantly taken from
\cite{Anderson2014} and \cite{Faran2014b}, but also included are the
historical Type~IIL SN~1979C \citep{Barbon1982} ($s50_V = 1.5$ mag)
and SN~1980K \citep{Barbon1982} ($s50_V = 2.1$ mag).  Each of the
SNe~IIL plotted in Figure~\ref{fig:lcV}
show a linear decay up until $\sim$80--120 days after explosion, followed by
a steep and rapid decline prior to reaching a secondary linear decline
phase powered by radioactive decay.  All SNe~IIL that have been
followed for more that 80 days from discovery show the drop in
magnitude that is characteristic of SNe~IIP. Based on  observations 
of several SNe~IIL,  \cite{Anderson2014} noted a similar luminosity drop, 
however, their photometric coverage is not as dense as that presented 
here (particularly at late phases), so a drop could not be robustly 
demonstrated for most of their events.
The light curve coverage of SN~2013by is such that it 
affords the best coverage of a late-time luminosity drop for a SN~IIL.
We also performed an extensive literature search for SNe~IIL that do 
not show this light curve drop, and found that only SN~1979C may 
have been one such object. Actually also SN~1979C, as suggested by 
\cite{Anderson2014}, may show a light curve drop around 50 days after 
explosion, though, if there, the drop would have occurred quite early 
and less pronounced than all the other cases.
It is also worth mentioning (see Figure~\ref{fig:lcV}) that the light curve drop to the radioactive
decay tail in SNe~IIL occurs at $\sim $80--100 days (versus $\sim $
100-140 days for SNe~IIP, confirming the correlation between the slope
of the plateau and plateau length previously reported by \cite{Anderson2014}.

\label{sec:data}

\section{Spectroscopy}
\label{sec:spectra}
\begin{figure*}
\begin{center}
  \includegraphics[width=16cm,height=10cm]{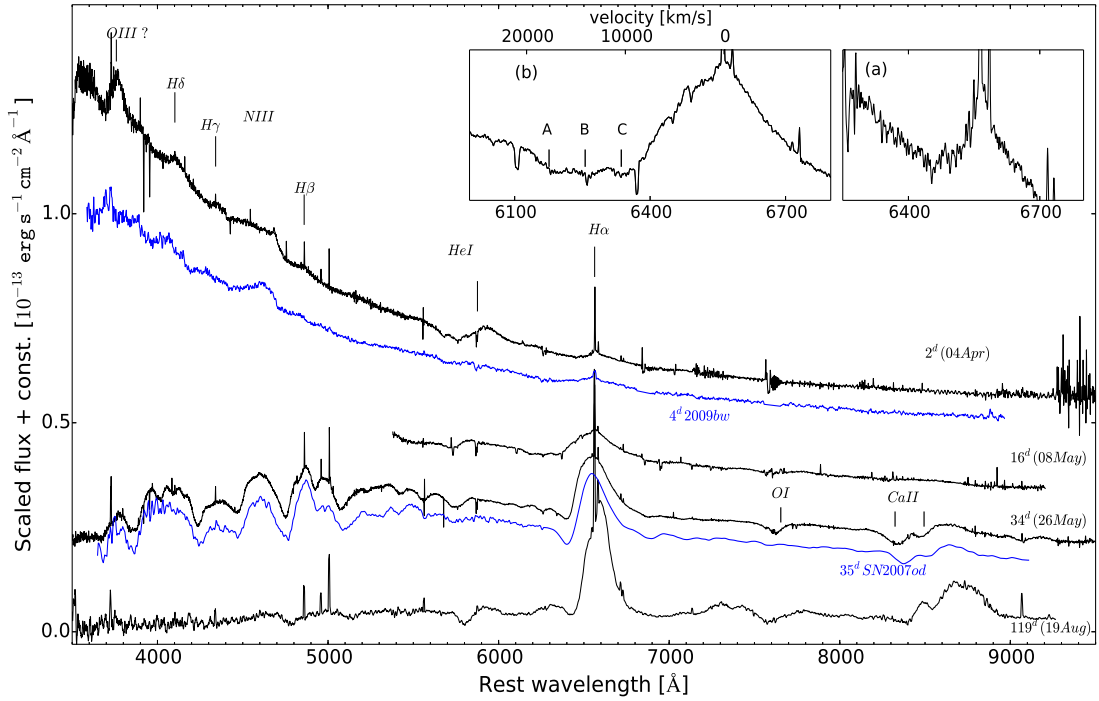}
  \caption{Visual-wavelengths spectra of SN~2013by. Inset (a) and (b) show the region 
blue-wards of \Ha{}, and highlight absorption features labeled as A, B, and C.} 
  \label{fig:spectra}
\end{center}
\end{figure*}

\begin{figure*}
\begin{center}
  \includegraphics[width=16cm,height=10cm]{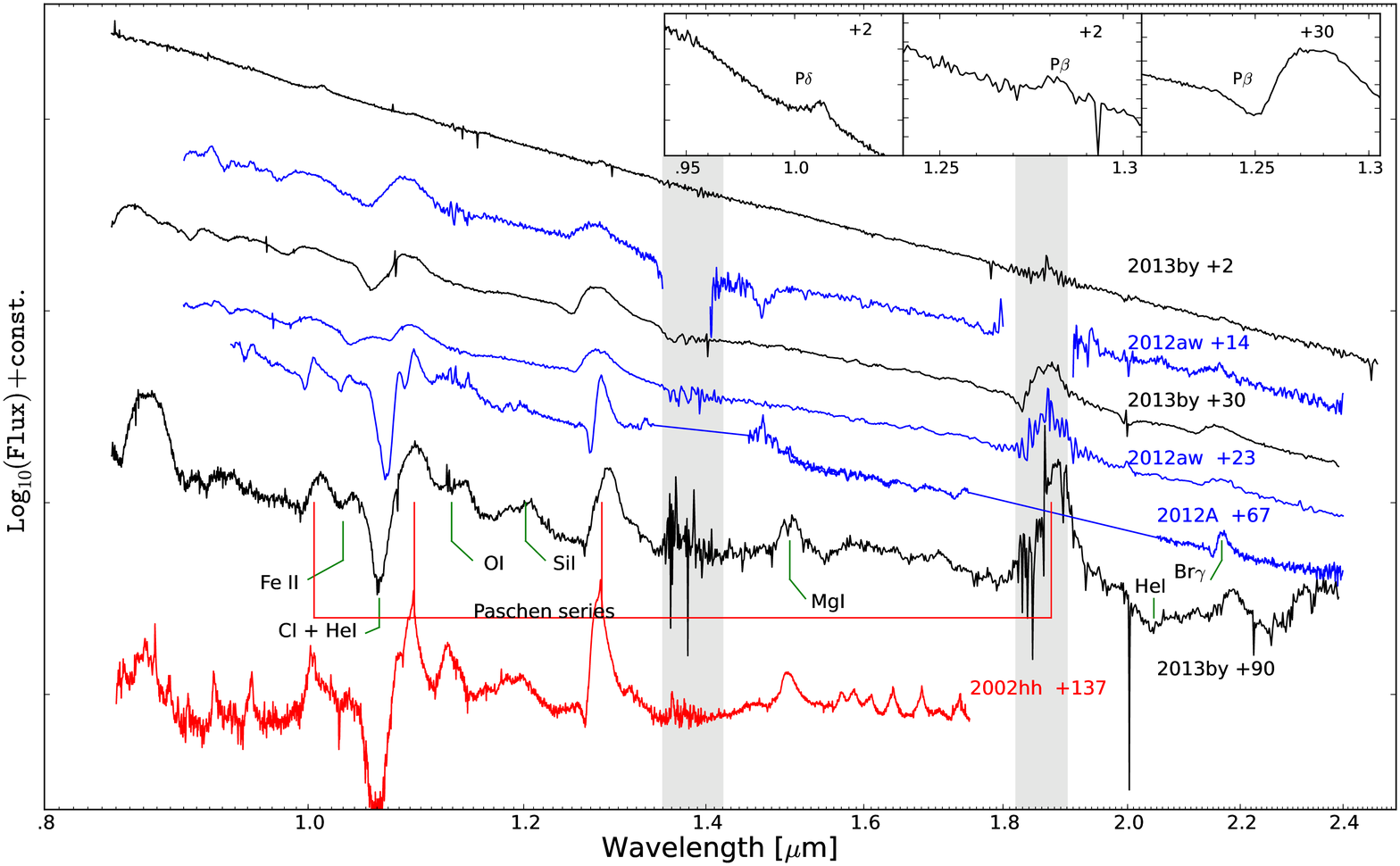}
  \caption{Infrared spectra of SN~2013by and  several Type  IIP SNe from the literature. 
  The three inset panels show the asymmetric profile of Paschen lines in the spectra at 
  2 days ($P\beta{}$ and $P\delta{}$) and 30 days ($P\beta{}$) after the shock breakout.  }
  \label{fig:infraredspectra}
\end{center}
\end{figure*}

The visual-wavelength spectra of SN~2013by are plotted in
Figure~\ref{fig:spectra}, while the NIR spectra are shown in
Figure~\ref{fig:infraredspectra}.  Also included in
Figure~\ref{fig:spectra} are comparison spectra of SN~2009bw and
SN~2007od, which are found by the spectral classification tool GELATO
\citep{Harutyunyan2008} to best match the spectra of SN~2013by  
at 16 and 34 days after shock breakout.
For the first spectrum of SN~2013by, GELATO find as best fit 
SN~1998S and several other SNe~IIn. However the fit are poor, 
probably for the lack of SNe~IIL spectra obtained at early phases. 
The spectrum of SN~2009bw obtained 4 days after shock breakout
shows several similarities to SN~2013by, although the latter exhibits 
conspicuous \HeI{} $\lambda$5876 (see Fig. \ref{fig:spectra}).

The first optical spectrum exhibits a blue continuum with the Balmer and the 
Paschen series clearly detected.  Both \HeI{} and \OIII{} are detected in the first optical 
spectrum of SN~2013by. As the SN evolves the Balmer lines become 
more prominent  and typical features of SNe~II emerge including: \CaII{} (H$\&$K
$\lambda\lambda$3934, 3968 \AA{}), Fe II (especially lines
$\lambda\lambda$4924, 5018 and 5169 \AA{}), \TiII{} (with many
multiplets below 5400\AA{}), \OI{} $\lambda$7774 \AA{} and the \CaII{}
infrared triplet $\lambda\lambda$8498, 8542 and 8662 \AA{}).  
The spectra at 16 and 34 days after explosion show 
several absorption features blueshifted with respect to photospheric \Ha{} 
(marked as A, B and C in Figure~\ref{fig:spectra}b). While  absorption feature 
A is consistent with \SiII{}, absorption features  B and C are probably due 
to \Ha{} at 15000~\kms{} and 8000~\kms{}, respectively.  
As mentioned in Section~\ref{sec:lightcurve}, these high-velocity features have 
also been identified in SN~2009bw and SN~2007od  
\citep{2011MNRAS.417..261I,2012MNRAS.422.1122I},
and interpreted as having an origin related to the interaction between rapidly expanding SN 
ejecta and circumstellar material (CSM).

Besides the overall similarity with SN~2007od, SN~2013by does show one
clear difference: the \OI{} $\lambda$7774 absorption line in the
spectrum of SN~2013by at 34 days after the explosion is quite
prominent, while it is almost absent in the spectrum of SN~2007od.
\cite{Faran2014a} have recently shown that oxygen seems to be more
pronounced in SNe IIL than in SNe IIP and it can be interpreted as a
sign of a more massive progenitor. However, oxygen should be used
carefully as a progenitor tracer for SNe
IIP. \cite{2012MNRAS.420.3451M} and \cite{Jerkstrand2012} have shown
that part of the oxygen visible in Type IIP SNe is synthesized soon
after the explosion, while the rest is primordial oxygen that is mixed
in the envelope during the evolution of the progenitor. Disentangling
these contributions is not easy.

\begin{figure}
\begin{center}
  \includegraphics[width=8.cm,height=6cm]{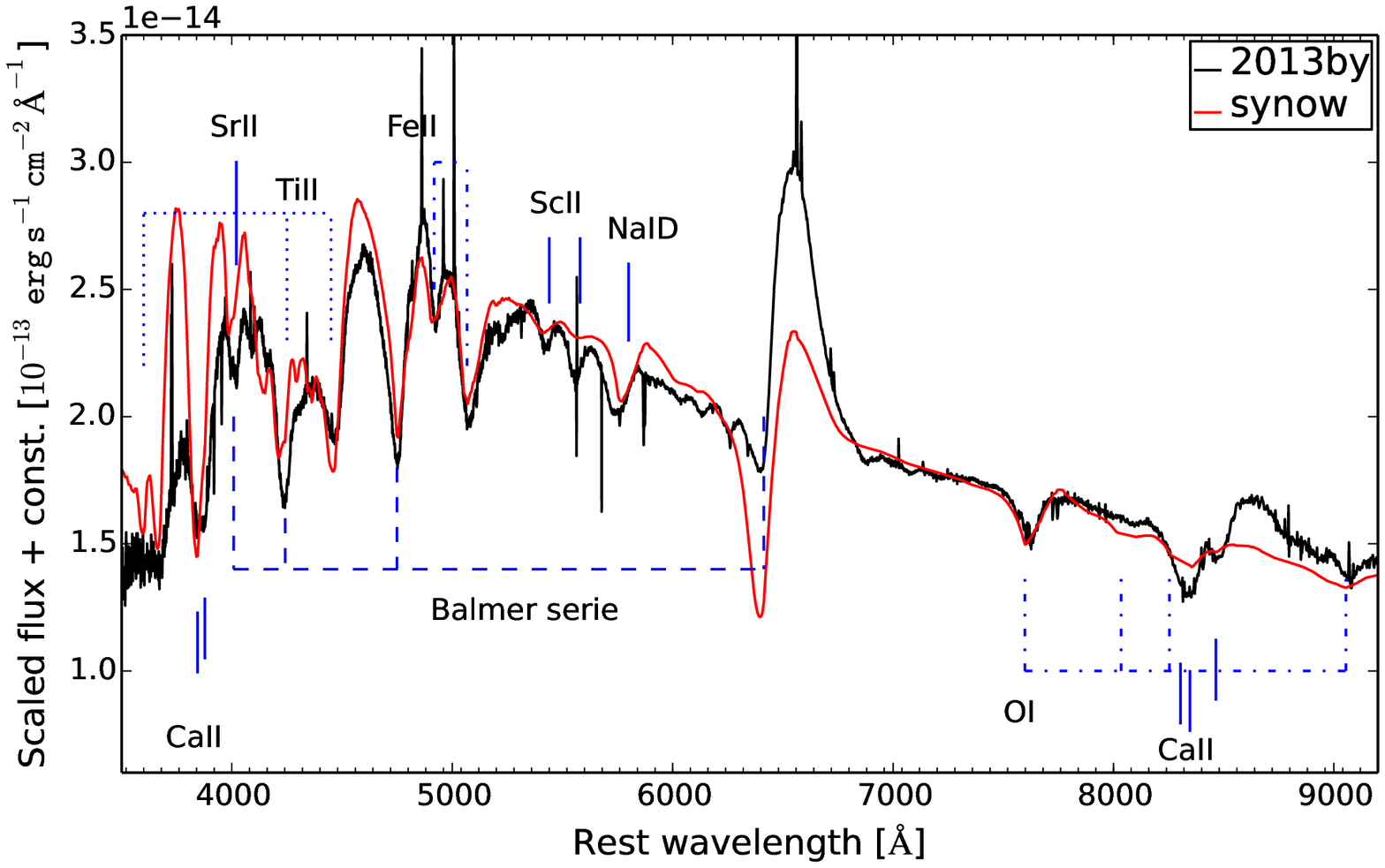}
  \caption{$Synow$ fit of the spectrum of SN~2013by at 34 days after the estimate shock breakout.}
  \label{fig:synow13by}
\end{center}
\end{figure}

A comparison of the observed visual-wavelength spectrum at 34 days after
the estimate shock breakout to our {\sc synow} fit in Figure~\ref{fig:synow13by}.
The spectrum has been reproduced with a black body temperature of 7200 K and 
a photospheric velocity of 7000 \kms{}. The following ions has been used 
in the synow spectrum: $HI$, \SrII{}, \ScII{}, \CaII{}, \OI{}, \FeII{}, \NaI{}, 
\TiII{}, and \SiII{}. The {\sc synow} spectrum well reproduces the observed 
spectrum, except for the ratio of the \Ha{} / \Hb{}. 
This is a well known problem related to the fact that {\sc synow} is 
based on the underlying assumption of local thermodynamical equilibrium \citep{Dessart2010}. 
 
Plotted in Figure~\ref{fig:infraredspectra} are the 3 FIRE NIR spectra
of SN~2013by compared to the spectra of other SNe~II, including SNe
2002hh \citep{Pozzo2006}, 2012A \citep{Tomasella2013} and 2012aw
\citep{Dall'Ora2014}.  Beside the asymmetric profile of Hydrogen
lines, also visible in the optical spectra, the NIR spectra of
SN~2013by are similar to those of other Type II SNe.  Of particular
interest is the last ($+$90d) FIRE spectrum of SN~2013by, which, to
our knowledge, is one of only a handful of NIR late-time spectra
published to date for a Type II SN, and the first of a Type IIL
SN. \cite{Pozzo2006} published a large set of infrared spectra of
SN~2002hh at late time.

SN~2002hh with a $s50_V$ = 0.27 mag/50 days is a Type IIP SN. The
first of their spectra (+137d) is shown in the comparison, and several
of their line identifications have provided us with guidelines for our
analysis of SN~2013by.  Beside the Brackett series which is clearly
visible in the spectrum of SN~2002hh in the range 16000-18000 \AA{},
most of the other lines are also visible in the last spectrum of
SN~2013by. This includes, e.g., \FeII{}, \OI{}, \MgI{}, \HeI{} and
\SiI{}.

The presence of \OI{} at $\lambda$11290 \AA{} confirms the \OI{}
identification at optical wavelengths. As reported by
\cite{Pozzo2006}, this line is a Bowen resonance fluorescence line
which requires the presence of microscopic mixing of hydrogen and
oxygen in the ejecta to be excited. \HeI{} is also clearly visible at
$\lambda$10830 \AA{} and $\lambda$20580 \AA{}.  The presence of \HeI{}
at this phase can be explained only if iron-group elements are mixed
in the ejecta.  The presence of \OI{} and \HeI{} confirm that
significant mixing has occurred within the ejecta.  \MgI{}
$\lambda$15030 \AA{}, is also clearly detected, while no sign of
[\SiI] $\lambda$16068 \AA{} is apparent.

\section{$^{56}$Ni in SNe IIP/IIL}
\label{sec:nickel}

Recent studies of SNe~II have shown that the Type~IIL objects are on
average more luminous at peak brightness than Type~IIP objects
($M_{IIL}=-17.44$ and $M_{IIP}=-15.66$, \citealt{Li2011}; see also
\citealt{Anderson2014,Faran2014a} and \citealt{Sanders2014}).  This is
consistent with a scenario where the envelope of SNe~IIL progenitors
retains a much smaller amount of hydrogen than their SNe~IIP
counterparts.  A massive hydrogen-rich envelope causes a slower
release of energy and a fainter maximum luminosity due to its
ionization and expansion during the SN explosion \citep{Patat1994}.
A contribution to the extra luminosity of SNe IIL may also come 
from CSM interaction (see Section~\ref{sec:interaction}).

An alternative explanation for the higher luminosity observed in
SNe~IIL is a larger amount of $^{56}$Ni synthesized during the
explosion.  The most appropriate way to estimate the $^{56}$Ni content
is by observing the supernova after $\sim$100 days from explosion,
when the $^{56}$Co decay becomes the dominate source of energy that
powers the broad-band emission.  Unfortunately SNe~IIL decline 4--5
magnitudes from peak and are often too faint to be observed at this
phase.  However, we were able to follow SN~2013by, and a few other
SNe~IIL for more than 100 days, recovering both the drop from the
plateau typical of SNe~IIP and the subsequent fall of the light curve
onto the radioactive tail.

Using SN~1987A as reference, we estimated the amount of $^{56}$Ni
produced during the explosion using the method of
\cite{Spiro2014}. This method consists of comparing the
pseudo-bolometric light curve of these objects with the
pseudo-bolometric light curve of SN~1987A (integrated in the same
bands) as soon as the SN fall onto the radioactive tail, such that

\begin{equation}
M_{SN}(Ni)= 0.075 \times  \frac{L_{SN}}{{L_{87A}}} M_{\odot}.
\end{equation}

\begin{figure}
\begin{center}
  \includegraphics[width=8.5cm,height=7cm]{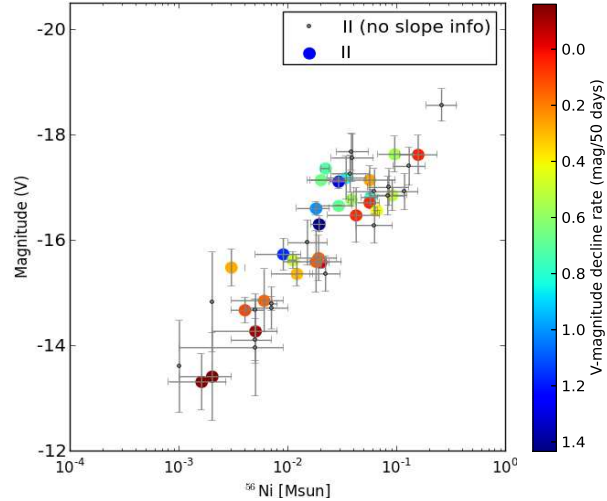}
  \caption{Absolute $V$-band magnitude at 50 days from explosion versus 
  $^{56}$Ni mass for a sample of  SNe~II.  SNe~IIL (blue points) cluster on the 
  top-left  side of the relation. The data are from \protect\cite{Hamuy2003}, \protect\cite{Spiro2014} and this work.}
  \label{fig:nickel}
\end{center}
\end{figure}

Armed with this methodology, we measure the $^{56}$Ni mass for 8
SNe~II (see Table~\ref{tabsummary}, recently published or with work in
preparation). \cite{Hamuy2003} and \cite{Spiro2014} presented a
similar analysis for a sample of 27 and 17 SNe~II, respectively.
Figure~\ref{fig:nickel} shows the relation between $^{56}$Ni mass and
the absolute $V$-band magnitude at 50 days from explosion for the 44
SNe (\citealt{Hamuy2003} and \citealt{Spiro2014}) and the 8 SNe from
Table~\ref{tabsummary}.  For a sub-sample of these SNe, we were able
to measure the $s50_V$ parameter. These SNe have been plotted in
Figure~\ref{fig:nickel} with different colors depending on their
$s50_V$ values (bluer points for larger slopes, red points for lower
slopes).

\cite{Hamuy2003} have shown a clear relation between $^{56}$Ni mass
and the absolute $V$-band magnitude (at 50 days from explosion).
\cite{Spiro2014} confirmed that faint SNe~IIP also follow the same
relation.  Even though the number of SNe~IIL with a $^{56}$Ni estimate
from the radioactive tail is small, these objects seem to follow the
same relations. However, part of the scatter in this relation seems to
be related to the $s50_{V}$ parameter.  Blue points (SNe IIL) sit slightly 
brighter than the general trend for a given $^{56}$Ni mass.

\section{SNe IIL and CSM interaction}
\label{sec:interaction}

We have shown in section \ref{sec:lightcurve} and section \ref{sec:spectra} 
that, SN~2013by is similar to the class of \emph{moderately interacting} Type II SNe.
X-ray emission has also been detected for SN~2013by with $Swift$ 
\citep{Margutti2013}. They measure a 0.3-1.0 keV count-rate of 2.1 $\pm$ 0.7 cps, 
that assuming a simple  power-law spectra model with photon index Gamma =2, 
translate in a flux of  1.1e-13 erg/s/cm2 (.3-10 keV).

This raises several questions: Are all Type IIL SNe interacting with CSM? Do all 
\emph{moderately interacting}  Type II SNe decline like SNe~IIL? It is not 
straight forward to generalize, but if SNe~IIL are coming from progenitors 
that lost most of their hydrogen envelope during pre-SN evolution, 
it is more likely that these SNe (more than Type IIP SNe) will show CSM interaction.
Unfortunately, for most of the objects we do not have enough 
information to answer these questions. It is indeed difficult to clearly 
separate between interacting and not interactive  SNe~II.

\begin{figure*}
\begin{center}
  \includegraphics[width=12cm,height=7cm]{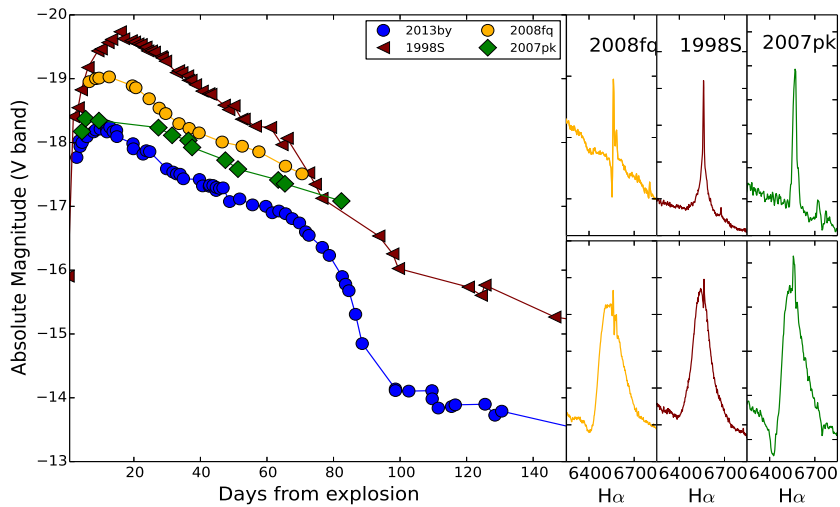}
  \caption{{\it (left panel)} $V$-band absolute magnitude light curve of SNe~II that show at early phases signs of interaction with CSM; {\it (right panel)} \Ha{} profiles soon after discovery (top) and at the end of the plateau phase (bottom) for three SNe~IIP/L-IIn.}
  \label{fig:interact}
\end{center}
\end{figure*}

At very early phases SNe IIP/IIL may be very similar to SNe IIn since 
the SN ejecta has not yet had time to reach and shock the CSM. 
SN~2008fq is a clear example of a SN~II that shows sign of 
interaction at very early phases.  It has been considered to 
be a SN~IIn by \cite{Taddia2013}
because of the clear detection of narrow lines in its spectra before
maximum, while \cite{Faran2014a} include this SN in their sample of
SNe IIL since no narrow lines are visible after maximum light.
Photometrically SN~2008fq and SN~2013by 
have similar slopes after maximum even though SN~2008fq 
is one magnitude brighter than SN~2013by (see right panel of Figure 
\ref{fig:13byIIPIIL} and Figure \ref{fig:interact}). Unfortunately
both light curves from \cite{Faran2014a} and \cite{Taddia2013} 
stop at $\sim$ 60 days, before the drop from the plateau would 
have occurred. Plotted in Figure~\ref{fig:interact}  (left panel) 
are the absolute $V$-band magnitude light curve of SN2008fq and 
three other SNe that (right, top panel) show signs of interaction with 
CSM at early phases.
 
\section{Conclusions}
\label{sec:discussion}

We have presented UV and optical broad-band photometry for the
Type~IIL SN~2013by ranging up to 150 days after explosion.  Our
extended and dense photometric coverage confirms that all SNe~IIL 
show a similar drop in the light curve down to the
radioactive tail as is seen in SNe~IIP, but at an earlier stage. If
SNe~IIL are followed for more than 80-100 days, they show that SNe IIP
and IIL share similar underling physics, supporting the idea that the
separation in two classes is purely nominal. Only a handful of objects
that decline as fast as SNe~IIL do not show the drop from the plateau,
suggesting that their light curves may be powered (also) by a
different source of energy (different than recombination).  We suggest
that the drop from the plateau (instead of the slope) should be use as
a more physical parameter to distinguish different types of SNe~II.

We have also presented visual-wavelength and NIR spectra of SN~2013by,
and have made a detailed comparison to similar data of other SNe~II.
The visual-wavelength spectra suggest that SN~2013by has experienced a
moderate amount of interaction between its rapidly expanding ejecta
and its CSM for more than one month after explosion.  Several SNe~II
show evidence of interaction with the CSM at early phases.  Most of
these objects are moderately luminous and they show a fast $V$ light
curve decline after maximum similar to SNe~IIL.  A late ($+$90d
relative to peak) NIR spectrum of SN~2013by exhibits similarities to a
NIR spectrum of the Type IIP SN~2002hh.  This comparison strengthens
the similarity between Type IIP and IIL SNe suggesting that strong
mixing occurs in the progenitors of both varieties.
 
We also investigate whether or not SNe~IIL are on average more
luminous than SNe~IIP, and if so, if this is related to the amount of
$^{56}$Ni synthesized during the explosion.  We use the magnitude
vs. $^{56}$Ni relations introduced by \citet{Hamuy2003} and
\citet{Spiro2014}, adding as extra information the slope in $V$-band
light curve ($s50_V$).  We find that SNe~IIL broadly follow these
relations, but for a similar amount of $^{56}$Ni produced during the
explosion, SNe~IIL are on average brighter than SNe~IIP.  This is in
agreement with the idea that SNe II with a larger ejected mass, have a
slower release of energy and a fainter maximum luminosity due to its
ionization and expansion during the SN explosion.

\begin{table*}
\caption{Main parameters for Type IIL  and IIP SNe (added to the previous works)}
\label{tabsummary}
\begin{tabular}{lllllllll}
\hline
Supernova   & Nickel  & $M_{V}^{a}$ & $M_{V50}^{b}$ & Distance            & $E(B-V)_{host}$ & $E(B-V)_{MW}^{d}$ & Explosion & Ref$^{e}$.\\
                   &             &                    &                         & modulus$^{c}$ &                          &                               & epoch      &        \\
\hline
SN~2003hn     & 0.038 (002)   &    $-$17.40 (14)  &  $-$16.78 (03) & 32.25  (12)    &  0.173   & 0.014  & 2452870.0   (4)       &  1 \\
SN~2009kr      & 0.009 (004)   &   $-$16.82 (30)  &  $-$15.74 (08) & 32.09   (50)    &  0.0       & 0.077  & 2455140.5   (2)       &  2\\
SN~2013by     & 0.029 (005)   &    $-$18.21 (14)  &  $-$17.12 (11)  & 30.85 (15)    &  0.0       & 0.195  & 2456404.0   (2)       &  3\\
SN~2013ej      & 0.018 (006)   &    $-$17.27 (13)  &  $-$16.61 (10)  & 29.79 (02)    &  0.0       &  0.061 &  2456497.4 (2)        & 4\\
SN~2013fs      & 0.057 (006)   &   $-$17.71 (15)  &   $-$16.82 (04)  & 33.5   (15)    &  0.0       & 0.035  &  2456571.2 (1)       & 5\\
SN~2014G      & 0.019 (003)   &    $-$17.46 (15)  &  $-$16.32 (08)  & 31.83  (02)   &  0.0       & 0.012  &  2456668.3 (1)       & 6\\
SN~2012A      & 0.011  (001)  &    $-$16.28   (16) &  $-$15.63 (08)  & 29.96 (15)   &  0.012   & 0.024  & 2455932.5  (2)      & 7 \\
SN~2012aw    & 0.056  (013)  &    $-$16.92 (10)  &  $-$16.72 (10)   &  29.96 (02)  &  0.028   &  0.058 & 2456002.5  (1)      & 8 \\
\hline
\end{tabular}\\
$^a$ Absolute magnitude at maximum. 
$^b$ Absolute magnitude at 50 days after explosion. 
$^c$ From NED, corrected for Local-Group infall onto the Virgo cluster and assuming $H_0 = 73\pm5$ \kms\,Mpc$^{-1}$.
$^d$ \protect\cite{1998ApJ...500..525S}. 
$^e$ References: 1=\cite{Krisciunas2009};  2=\cite{2010ApJ...714L.280F}, \cite{Elias-Rosa2011}; 3= this work, 4= \cite{Valenti2014}; 5= Trematerra et al in prep ; 
6= Yaron et al in prep.; 7= \cite{Tomasella2013} ; 8= \cite{Dall'Ora2014}.\\
\end{table*}

\section*{Acknowledgements}
This material is based upon work supported by NSF under grants
AST--0306969, AST--0908886, AST--0607438, and AST--1008343.
M.D.S. and the CSP gratefully acknowledge generous support provided by
the Danish Agency for Science and Technology and Innovation realized
through a Sapere Aude Level 2 grant.  MG acknowledges support from
Joined Committee ESO and Government of Chile 2014 and the Ministry for
the Economy, Development, and Tourisms Programa Inicativa Cientifica
Milenio through grant IC 12009, awarded to The Millennium Institute of
Astrophysics (MAS) and Fondecyt Regular No. 1120601.  We are grateful
to Rubina Kotak who provided us spectra of SN~2002hh.  This paper
includes data gathered with the 6.5 meter Magellan Telescopes located
at Las Campanas Observatory, Chile.  This paper is based on
observations made with the following facilities: Swift, LCOGT.



\appendix

\section[]{Tables}

\begin{table*}
 \centering
  \begin{minipage}{140mm}
  \caption[3]{Journal of spectroscopic observations.}
  \label{tab1}  
  \begin{tabular}{@{}ccccccc@{}}
  \hline   
Date  &   Telescope &  JD  & Phase\footnote{Relative to the date of the estimate 
shock breakout (SN~2013by, JD =  2456405.0; SN~2013ej, JD = 2456497.45).} &
Instrument & Range & Resolution FWHM\footnote{FWHM of night-sky emission lines.} 
 \\
 &  & $-$2,400,000 (days) & (days) &  & (\AA) & (\AA) \\
 \hline
  2013 Apr  24  & Magellan  & 56406.88   &  1.8     & FIRE & 9000-25000 &    4 \\
  2013 Apr  24 &  ANU 2.3m  &     56407.00   &  2.0     & WifeS & 3200-10000 &   1 \\
  2013 May  08 &  ANU 2.3m  &     56421.00   &  16.0     & WifeS & 3200-10000 &   1 \\
  2013 May  22  & Magellan   &     56435.01   &  30.0     & FIRE & 9000-25000 &    4 \\
  2013 May  26 &  ANU 2.3m  &     56421.000   &  34.0     & WifeS & 3200-10000 &   1 \\
  2013 Jul  29  & Magellan  &  56503.01   &  98.0     & FIRE & 9000-25000 &    4 \\
  2013 May  26 &  ANU 2.3m  &     56524.000   &  119.0     & WifeS & 3200-10000 &   1 \\
\hline
\end{tabular}
\end{minipage}
\end{table*}

\begin{table*}
\centering
 \begin{minipage}{140mm}
\caption{Photometric Data (Complete table available in the online version of the paper. Errors include photometry errors and errors on the nightly zero point.)}
\scriptsize
\begin{tabular}{ccccc|ccccc}
\hline
Date & JD & mag$^{a}$ &  Filter & telescope$^b$ &Date & JD & mag &  Filter & telescope
\footnote{(a): Data has not been corrected for extinction; (b): {\em Swift} Telescope; CSP (Las Campanas, Chile); 1m0-08 (McDonald observatory, USA); 1m0-10, 1m0-12, 1m0-13 (Sutherland, South Africa), 1m0-04, 1m0-05, 1m0-09  (Cerro Tololo, Chile); 1m0-03, 1m0-11 (Siding Spring, Australia).} \\
\hline
2013-04-24 &  2456407.372 &  12.763 0.070 & $uw2$ &  $Swift$ & 2013-06-04 &  2456448.092 &  16.409 0.070 & $U$ &  $Swift$ \\ 
2013-04-30 &  2456413.176 &  14.084 0.073 & $uw2$ &  $Swift$ & 2013-06-21 &  2456464.719 &  17.533 0.095 & $U$ &  $Swift$ \\ 
2013-05-09 &  2456422.406 &  16.076 0.089 & $uw2$ &  $Swift$ & 2013-06-24 &  2456467.659 &  17.680 0.114 & $U$ &  $Swift$ \\ 
2013-05-10 &  2456423.475 &  16.314 0.092 & $uw2$ &  $Swift$ & 2013-06-29 &  2456472.541 &  17.914 0.094 & $U$ &  $Swift$ \\ 
2013-05-12 &  2456425.345 &  16.908 0.102 & $uw2$ &  $Swift$ & 2013-07-03 &  2456477.411 &  18.643 0.106 & $U$ &  $Swift$ \\ 
2013-05-23 &  2456435.884 &  18.169 0.103 & $uw2$ &  $Swift$ & 2013-07-07 &  2456481.348 &  17.990 0.105 & $U$ &  $Swift$ \\ 
2013-05-26 &  2456439.121 &  18.410 0.095 & $uw2$ &  $Swift$ & 2013-07-11 &  2456485.430 &  18.652 0.120 & $U$ &  $Swift$ \\ 
2013-05-29 &  2456441.826 &  18.778 0.109 & $uw2$ &  $Swift$ & 2013-04-24 &  2456407.370 &  13.753 0.047 & $B$ &  $Swift$ \\ 
2013-06-04 &  2456448.096 &  18.794 0.108 & $uw2$ &  $Swift$ & 2013-05-23 &  2456435.882 &  14.753 0.048 & $B$ &  $Swift$ \\ 
2013-06-21 &  2456464.722 &  19.259 0.119 & $uw2$ &  $Swift$ & 2013-05-26 &  2456439.119 &  14.966 0.047 & $B$ &  $Swift$ \\ 
2013-06-24 &  2456467.661 &  20.678 0.153 & $uw2$ &  $Swift$ & 2013-05-29 &  2456441.823 &  15.112 0.049 & $B$ &  $Swift$ \\
2013-06-29 &  2456472.545 &  19.620 0.114 & $uw2$ &  $Swift$ & 2013-06-04 &  2456448.093 &  15.332 0.050 & $B$ &  $Swift$ \\ 
2013-07-03 &  2456477.415 &  20.695 0.119 & $uw2$ &  $Swift$ & 2013-06-21 &  2456464.720 &  15.847 0.054 & $B$ &  $Swift$ \\ 
2013-07-07 &  2456481.351 &  20.720 0.131 & $uw2$ &  $Swift$ & 2013-06-24 &  2456467.660 &  16.088 0.062 & $B$ &  $Swift$ \\ 
2013-07-11 &  2456485.433 &  20.346 0.132 & $uw2$ &  $Swift$ & 2013-06-29 &  2456472.542 &  16.249 0.055 & $B$ &  $Swift$ \\ 
2013-04-24 &  2456407.377 &  12.707 0.059 & $um2$ &  $Swift$ & 2013-07-03 &  2456477.412 &  16.880 0.062 & $B$ &  $Swift$ \\ 
2013-05-09 &  2456422.411 &  15.909 0.088 & $um2$ &  $Swift$ & 2013-07-07 &  2456481.349 &  16.722 0.064 & $B$ &  $Swift$ \\ 
2013-05-10 &  2456423.480 &  15.984 0.089 & $um2$ &  $Swift$ & 2013-07-11 &  2456485.431 &  17.231 0.074 & $B$ &  $Swift$ \\ 
2013-05-23 &  2456435.888 &  18.566 0.142 & $um2$ &  $Swift$ & 2013-04-24 &  2456406.830 &  13.732 0.008 & $B$ &      CSP \\ 
2013-05-26 &  2456439.124 &  18.873 0.124 & $um2$ &  $Swift$ & 2013-04-25 &  2456407.850 &  13.653 0.006 & $B$ &      CSP \\ 
2013-05-29 &  2456441.830 &  18.792 0.138 & $um2$ &  $Swift$ & 2013-04-27 &  2456409.860 &  13.496 0.005 & $B$ &      CSP \\ 
2013-06-04 &  2456448.100 &  18.892 0.152 & $um2$ &  $Swift$ & 2013-04-29 &  2456411.920 &  13.468 0.005 & $B$ &      CSP \\ 
2013-06-21 &  2456464.726 &  19.709 0.163 & $um2$ &  $Swift$ & 2013-05-01 &  2456413.910 &  13.516 0.013 & $B$ &      CSP \\ 
2013-06-24 &  2456467.664 &  18.999 0.174 & $um2$ &  $Swift$ & 2013-05-03 &  2456415.900 &  13.559 0.005 & $B$ &      CSP \\ 
2013-06-29 &  2456472.550 &  19.318 0.155 & $um2$ &  $Swift$ & 2013-05-06 &  2456418.900 &  13.684 0.005 & $B$ &      CSP \\ 
2013-07-07 &  2456481.355 &  19.489 0.177 & $um2$ &  $Swift$ & 2013-05-11 &  2456423.850 &  13.939 0.006 & $B$ &      CSP \\ 
2013-07-11 &  2456485.437 &  19.502 0.162 & $um2$ &  $Swift$ & 2013-05-14 &  2456426.770 &  14.118 0.005 & $B$ &      CSP \\ 
2013-04-24 &  2456407.367 &  12.566 0.057 & $uw1$ &  $Swift$ & 2013-05-21 &  2456433.840 &  14.660 0.006 & $B$ &      CSP \\ 
2013-04-30 &  2456413.183 &  13.206 0.058 & $uw1$ &  $Swift$ & 2013-05-23 &  2456435.750 &  14.784 0.006 & $B$ &      CSP \\ 
2013-05-23 &  2456435.879 &  16.837 0.087 & $uw1$ &  $Swift$ & 2013-05-24 &  2456436.850 &  14.863 0.005 & $B$ &      CSP \\ 
2013-05-26 &  2456439.117 &  17.034 0.080 & $uw1$ &  $Swift$ & 2013-05-25 &  2456437.870 &  14.930 0.006 & $B$ &      CSP \\ 
2013-05-29 &  2456441.821 &  17.167 0.091 & $uw1$ &  $Swift$ & 2013-05-26 &  2456438.870 &  14.984 0.005 & $B$ &      CSP \\ 
2013-06-04 &  2456448.090 &  17.437 0.096 & $uw1$ &  $Swift$ & 2013-06-01 &  2456444.678 &  15.210 0.006 & $B$ &      CSP \\ 
2013-06-21 &  2456464.718 &  18.182 0.119 & $uw1$ &  $Swift$ & 2013-06-03 &  2456446.685 &  15.326 0.011 & $B$ &      CSP \\ 
2013-06-24 &  2456467.658 &  18.344 0.144 & $uw1$ &  $Swift$ & 2013-06-05 &  2456448.732 &  15.381 0.009 & $B$ &      CSP \\ 
2013-06-29 &  2456472.539 &  18.167 0.108 & $uw1$ &  $Swift$ & 2013-04-25 &  2456407.560 &  13.563 0.021 & $B$ &   1m0-10 \\ 
2013-07-03 &  2456477.409 &  18.769 0.117 & $uw1$ &  $Swift$ & 2013-04-25 &  2456408.459 &  13.507 0.006 & $B$ &   1m0-10 \\ 
2013-07-07 &  2456481.347 &  18.790 0.132 & $uw1$ &  $Swift$ & 2013-04-27 &  2456410.459 &  13.436 0.011 & $B$ &   1m0-10 \\ 
2013-07-11 &  2456485.428 &  18.931 0.138 & $uw1$ &  $Swift$ & 2013-04-28 &  2456411.459 &  13.482 0.017 & $B$ &   1m0-10 \\ 
2013-04-24 &  2456406.830 &  13.638 0.018 & $u$ &      CSP & 2013-04-30 &  2456412.500 &  13.411 0.019 & $B$ &   1m0-10 \\ 
2013-04-25 &  2456407.850 &  13.561 0.009 & $u$ &      CSP & 2013-05-03 &  2456415.845 &  13.586 0.007 & $B$ &    1m0-5 \\ 
2013-04-27 &  2456409.850 &  13.520 0.008 & $u$ &      CSP & 2013-05-04 &  2456416.801 &  13.632 0.007 & $B$ &    1m0-5 \\ 
2013-04-29 &  2456411.920 &  13.531 0.007 & $u$ &      CSP & 2013-05-05 &  2456417.779 &  13.613 0.012 & $B$ &    1m0-4 \\ 
2013-05-01 &  2456413.900 &  13.601 0.011 & $u$ &      CSP & 2013-05-06 &  2456418.778 &  13.667 0.007 & $B$ &    1m0-4 \\ 
2013-05-03 &  2456415.910 &  13.726 0.006 & $u$ &      CSP & 2013-05-11 &  2456423.738 &  13.919 0.007 & $B$ &    1m0-9 \\ 
2013-05-06 &  2456418.910 &  13.956 0.008 & $u$ &      CSP & 2013-05-15 &  2456427.738 &  14.197 0.015 & $B$ &    1m0-9 \\ 
2013-05-11 &  2456423.850 &  14.540 0.008 & $u$ &      CSP & 2013-05-16 &  2456428.749 &  14.242 0.009 & $B$ &    1m0-9 \\ 
2013-05-14 &  2456426.780 &  14.931 0.009 & $u$ &      CSP & 2013-05-31 &  2456443.706 &  15.113 0.005 & $B$ &    1m0-9 \\ 
2013-05-21 &  2456433.860 &  15.873 0.014 & $u$ &      CSP & 2013-06-03 &  2456446.786 &  15.219 0.029 & $B$ &    1m0-4 \\ 
2013-05-23 &  2456435.770 &  16.074 0.011 & $u$ &      CSP & 2013-06-04 &  2456447.786 &  15.241 0.011 & $B$ &    1m0-4 \\ 
2013-05-24 &  2456436.870 &  16.154 0.012 & $u$ &      CSP & 2013-06-05 &  2456448.786 &  15.298 0.002 & $B$ &    1m0-4 \\ 
2013-05-25 &  2456437.870 &  16.354 0.013 & $u$ &      CSP & 2013-06-06 &  2456449.785 &  15.315 0.009 & $B$ &    1m0-4 \\ 
2013-05-26 &  2456438.860 &  16.414 0.012 & $u$ &      CSP & 2013-06-07 &  2456450.788 &  15.331 0.011 & $B$ &    1m0-4 \\ 
2013-06-01 &  2456444.673 &  16.803 0.012 & $u$ &      CSP & 2013-06-09 &  2456452.787 &  15.417 0.011 & $B$ &    1m0-4 \\ 
2013-06-03 &  2456446.694 &  17.029 0.021 & $u$ &      CSP & 2013-06-12 &  2456455.786 &  15.522 0.000 & $B$ &    1m0-4 \\ 
2013-06-05 &  2456448.742 &  17.186 0.025 & $u$ &      CSP & 2013-06-14 &  2456457.791 &  15.588 0.019 & $B$ &    1m0-4 \\ 
2013-04-24 &  2456407.369 &  12.521 0.053 & $U$ &  $Swift$ & 2013-06-16 &  2456459.658 &  15.674 0.010 & $B$ &    1m0-9 \\ 
2013-05-23 &  2456435.881 &  15.392 0.060 & $U$ &  $Swift$ & 2013-06-18 &  2456461.670 &  15.648 0.008 & $B$ &    1m0-4 \\ 
2013-05-26 &  2456439.118 &  15.780 0.059 & $U$ &  $Swift$ & 2013-06-20 &  2456463.657 &  15.789 0.016 & $B$ &    1m0-4 \\ 
2013-05-29 &  2456441.822 &  16.026 0.066 & $U$ &  $Swift$ & 2013-06-22 &  2456465.592 &  15.866 0.004 & $B$ &    1m0-4 \\ 
 \hline
\end{tabular}
\label{tab2}
\end{minipage}
\end{table*}

\begin{table*}
    \centering
    \begin{minipage}{140mm}
    \caption{Photometric Data}
    \begin{tabular}{ccccc|ccccc}
    \hline 
Date &JD &mag &Filter &telescope &Date &JD &mag &Filter &telescope \footnote{{\em Swift} Telescope; CSP (Las Campanas, Chile); 1m0-08 (McDonald observatory, USA); 1m0-10, 1m0-12, 1m0-13 (Sutherland, South Africa), 1m0-04, 1m0-05, 1m0-09  (Cerro Tololo, Chile); 1m0-03, 1m0-11 (Siding Spring, Australia)}\\ 
 \hline 
2013-06-24 &  2456467.592 &  15.941 0.014 & $B$ &    1m0-4 & 2013-07-19 &  2456492.626 &  17.005 0.003 & $g$ &    1m0-4 \\ 
2013-06-26 &  2456469.592 &  16.005 0.009 & $B$ &    1m0-4 & 2013-07-29 &  2456502.622 &  17.794 0.041 & $g$ &    1m0-4 \\ 
2013-06-28 &  2456471.592 &  16.100 0.011 & $B$ &    1m0-4 & 2013-07-29 &  2456502.626 &  17.740 0.050 & $g$ &    1m0-9 \\ 
2013-06-30 &  2456473.758 &  16.248 0.023 & $B$ &    1m0-5 & 2013-07-31 &  2456504.692 &  17.929 0.031 & $g$ &    1m0-9 \\ 
2013-07-02 &  2456475.616 &  16.300 0.004 & $B$ &    1m0-4 & 2013-08-02 &  2456506.625 &  17.833 0.025 & $g$ &    1m0-5 \\ 
2013-07-03 &  2456476.616 &  16.376 0.005 & $B$ &    1m0-4 & 2013-08-09 &  2456513.600 &  17.901 0.022 & $g$ &    1m0-9 \\ 
2013-07-07 &  2456480.616 &  16.637 0.002 & $B$ &    1m0-4 & 2013-08-09 &  2456513.662 &  17.975 0.023 & $g$ &    1m0-4 \\ 
2013-07-09 &  2456482.745 &  16.830 0.007 & $B$ &    1m0-5 & 2013-08-10 &  2456515.492 &  18.133 0.227 & $g$ &    1m0-4 \\ 
2013-07-13 &  2456486.612 &  17.186 0.012 & $B$ &    1m0-4 & 2013-08-14 &  2456519.480 &  18.158 0.003 & $g$ &    1m0-9 \\ 
2013-07-14 &  2456487.612 &  17.363 0.002 & $B$ &    1m0-4 & 2013-08-16 &  2456520.594 &  18.133 0.023 & $g$ &    1m0-5 \\ 
2013-07-15 &  2456488.612 &  17.445 0.032 & $B$ &    1m0-4 & 2013-08-25 &  2456529.548 &  18.000 0.017 & $g$ &    1m0-9 \\ 
2013-07-17 &  2456490.612 &  17.754 0.007 & $B$ &    1m0-4 & 2013-08-28 &  2456532.523 &  18.134 0.008 & $g$ &    1m0-5 \\ 
2013-07-19 &  2456492.620 &  18.213 0.015 & $B$ &    1m0-4 & 2013-08-30 &  2456534.553 &  18.269 0.012 & $g$ &    1m0-4 \\ 
2013-07-27 &  2456500.537 &  18.692 0.200 & $B$ &    1m0-4 & 2013-09-21 &  2456556.514 &  18.263 0.019 & $g$ &    1m0-4 \\ 
2013-07-29 &  2456502.612 &  18.745 0.126 & $B$ &    1m0-4 & 2013-04-24 &  2456406.820 &  13.649 0.009 & $g$ &      CSP \\ 
2013-07-29 &  2456502.617 &  18.927 0.077 & $B$ &    1m0-9 & 2013-04-25 &  2456407.850 &  13.539 0.006 & $g$ &      CSP \\ 
2013-08-02 &  2456506.616 &  18.618 0.044 & $B$ &    1m0-5 & 2013-04-27 &  2456409.850 &  13.387 0.005 & $g$ &      CSP \\ 
2013-08-09 &  2456513.586 &  18.616 0.006 & $B$ &    1m0-9 & 2013-04-29 &  2456411.920 &  13.351 0.005 & $g$ &      CSP \\ 
2013-08-09 &  2456513.650 &  18.755 0.019 & $B$ &    1m0-4 & 2013-05-01 &  2456413.900 &  13.363 0.010 & $g$ &      CSP \\ 
2013-08-10 &  2456515.480 &  19.015 0.031 & $B$ &    1m0-4 & 2013-05-03 &  2456415.900 &  13.399 0.005 & $g$ &      CSP \\ 
2013-08-14 &  2456519.468 &  19.033 0.017 & $B$ &    1m0-9 & 2013-05-06 &  2456418.910 &  13.510 0.005 & $g$ &      CSP \\ 
2013-08-16 &  2456520.582 &  19.101 0.017 & $B$ &    1m0-5 & 2013-05-11 &  2456423.850 &  13.703 0.005 & $g$ &      CSP \\ 
2013-08-25 &  2456529.538 &  18.959 0.048 & $B$ &    1m0-9 & 2013-05-14 &  2456426.770 &  13.824 0.005 & $g$ &      CSP \\ 
2013-08-28 &  2456532.513 &  19.051 0.056 & $B$ &    1m0-5 & 2013-05-21 &  2456433.850 &  14.171 0.005 & $g$ &      CSP \\ 
2013-08-30 &  2456534.541 &  19.061 0.042 & $B$ &    1m0-4 & 2013-05-23 &  2456435.760 &  14.262 0.006 & $g$ &      CSP \\ 
2013-08-30 &  2456534.593 &  19.160 0.067 & $B$ &    1m0-9 & 2013-05-24 &  2456436.860 &  14.313 0.005 & $g$ &      CSP \\ 
2013-09-21 &  2456556.503 &  18.875 0.018 & $B$ &    1m0-4 & 2013-05-25 &  2456437.860 &  14.355 0.005 & $g$ &      CSP \\ 
2013-09-24 &  2456559.564 &  19.083 0.065 & $B$ &    1m0-4 & 2013-05-26 &  2456438.860 &  14.390 0.005 & $g$ &      CSP \\ 
2013-04-25 &  2456407.545 &  13.519 0.015 & $g$ &   1m0-10 & 2013-06-01 &  2456444.666 &  14.579 0.005 & $g$ &      CSP \\ 
2013-04-25 &  2456408.462 &  13.428 0.009 & $g$ &   1m0-10 & 2013-06-03 &  2456446.688 &  14.617 0.009 & $g$ &      CSP \\ 
2013-04-30 &  2456412.503 &  13.329 0.008 & $g$ &   1m0-10 & 2013-06-05 &  2456448.735 &  14.675 0.009 & $g$ &      CSP \\ 
2013-05-03 &  2456415.783 &  13.379 0.028 & $g$ &    1m0-4 & 2013-04-24 &  2456406.830 &  13.683 0.009 & $V$ &      CSP \\ 
2013-05-03 &  2456415.856 &  13.361 0.003 & $g$ &    1m0-5 & 2013-04-25 &  2456407.850 &  13.512 0.006 & $V$ &      CSP \\ 
2013-05-04 &  2456416.813 &  13.527 0.007 & $g$ &    1m0-5 & 2013-04-27 &  2456409.860 &  13.359 0.005 & $V$ &      CSP \\ 
2013-05-05 &  2456417.783 &  13.478 0.004 & $g$ &    1m0-4 & 2013-04-29 &  2456411.920 &  13.283 0.005 & $V$ &      CSP \\ 
2013-05-06 &  2456418.783 &  13.510 0.002 & $g$ &    1m0-4 & 2013-05-01 &  2456413.900 &  13.238 0.008 & $V$ &      CSP \\ 
2013-05-11 &  2456423.746 &  13.694 0.027 & $g$ &    1m0-9 & 2013-05-03 &  2456415.900 &  13.284 0.005 & $V$ &      CSP \\ 
2013-05-15 &  2456427.745 &  13.815 0.011 & $g$ &    1m0-9 & 2013-05-06 &  2456418.900 &  13.357 0.006 & $V$ &      CSP \\ 
2013-05-31 &  2456443.715 &  14.492 0.007 & $g$ &    1m0-9 & 2013-05-11 &  2456423.850 &  13.548 0.005 & $V$ &      CSP \\ 
2013-06-03 &  2456446.791 &  14.583 0.005 & $g$ &    1m0-4 & 2013-05-14 &  2456426.770 &  13.632 0.005 & $V$ &      CSP \\ 
2013-06-04 &  2456447.791 &  14.581 0.009 & $g$ &    1m0-4 & 2013-05-21 &  2456433.850 &  13.862 0.005 & $V$ &      CSP \\ 
2013-06-05 &  2456448.792 &  14.673 0.006 & $g$ &    1m0-4 & 2013-05-23 &  2456435.760 &  13.912 0.006 & $V$ &      CSP \\ 
2013-06-07 &  2456450.796 &  14.703 0.025 & $g$ &    1m0-4 & 2013-05-24 &  2456436.850 &  13.942 0.005 & $V$ &      CSP \\ 
2013-06-09 &  2456452.792 &  14.729 0.000 & $g$ &    1m0-4 & 2013-05-25 &  2456437.870 &  13.946 0.005 & $V$ &      CSP \\ 
2013-06-12 &  2456455.792 &  14.802 0.004 & $g$ &    1m0-4 & 2013-05-26 &  2456438.870 &  14.018 0.005 & $V$ &      CSP \\ 
2013-06-16 &  2456459.666 &  14.862 0.026 & $g$ &    1m0-9 & 2013-06-01 &  2456444.677 &  14.129 0.005 & $V$ &      CSP \\ 
2013-06-20 &  2456463.666 &  14.977 0.007 & $g$ &    1m0-4 & 2013-06-03 &  2456446.686 &  14.136 0.009 & $V$ &      CSP \\ 
2013-06-22 &  2456465.598 &  15.025 0.005 & $g$ &    1m0-4 & 2013-06-05 &  2456448.734 &  14.201 0.006 & $V$ &      CSP \\ 
2013-06-24 &  2456467.598 &  15.079 0.002 & $g$ &    1m0-4 & 2013-04-25 &  2456407.566 &  13.423 0.007 & $V$ &   1m0-10 \\ 
2013-06-26 &  2456469.598 &  15.129 0.013 & $g$ &    1m0-4 & 2013-04-25 &  2456408.461 &  13.449 0.004 & $V$ &   1m0-10 \\ 
2013-06-28 &  2456471.598 &  15.237 0.010 & $g$ &    1m0-4 & 2013-04-30 &  2456412.502 &  13.254 0.005 & $V$ &   1m0-10 \\ 
2013-06-30 &  2456473.763 &  15.317 0.021 & $g$ &    1m0-5 & 2013-05-03 &  2456415.781 &  13.191 0.024 & $V$ &    1m0-4 \\ 
2013-07-02 &  2456475.622 &  15.369 0.002 & $g$ &    1m0-4 & 2013-05-03 &  2456415.851 &  13.250 0.020 & $V$ &    1m0-5 \\ 
2013-07-03 &  2456476.622 &  15.464 0.009 & $g$ &    1m0-4 & 2013-05-04 &  2456416.809 &  13.209 0.004 & $V$ &    1m0-5 \\ 
2013-07-07 &  2456480.622 &  15.688 0.013 & $g$ &    1m0-4 & 2013-05-05 &  2456417.781 &  13.277 0.001 & $V$ &    1m0-4 \\ 
2013-07-09 &  2456482.751 &  15.831 0.002 & $g$ &    1m0-5 & 2013-05-06 &  2456418.781 &  13.261 0.010 & $V$ &    1m0-4 \\ 
2013-07-13 &  2456486.619 &  16.161 0.019 & $g$ &    1m0-4 & 2013-05-11 &  2456423.742 &  13.468 0.005 & $V$ &    1m0-9 \\ 
2013-07-14 &  2456487.618 &  16.272 0.005 & $g$ &    1m0-4 & 2013-05-15 &  2456427.742 &  13.578 0.010 & $V$ &    1m0-9 \\ 
2013-07-15 &  2456488.619 &  16.407 0.011 & $g$ &    1m0-4 & 2013-05-16 &  2456428.754 &  13.593 0.005 & $V$ &    1m0-9 \\ 
2013-07-17 &  2456490.622 &  16.523 0.038 & $g$ &    1m0-4 & 2013-05-31 &  2456443.712 &  14.029 0.008 & $V$ &    1m0-9 \\ 
\hline
    \end{tabular}
    \label{tab2}
    \end{minipage}
    \end{table*}

\begin{table*}
    \centering
    \begin{minipage}{140mm}
    \caption{Photometric Data}
    \begin{tabular}{ccccc|ccccc}
    \hline 
Date &JD &mag &Filter &telescope &Date &JD &mag &Filter &telescope \footnote{{\em Swift} Telescope; CSP (Las Campanas, Chile); 1m0-08 (McDonald observatory, USA); 1m0-10, 1m0-12, 1m0-13 (Sutherland, South Africa), 1m0-04, 1m0-05, 1m0-09  (Cerro Tololo, Chile); 1m0-03, 1m0-11 (Siding Spring, Australia)}\\ 
 \hline 
2013-06-03 &  2456446.788 &  14.116 0.002 & $V$ &    1m0-4 & 2013-06-28 &  2456471.601 &  14.105 0.011 & $r$ &    1m0-4 \\ 
2013-06-04 &  2456447.788 &  14.123 0.018 & $V$ &    1m0-4 & 2013-06-30 &  2456473.766 &  14.197 0.013 & $r$ &    1m0-5 \\ 
2013-06-05 &  2456448.789 &  14.152 0.008 & $V$ &    1m0-4 & 2013-07-02 &  2456475.625 &  14.225 0.001 & $r$ &    1m0-4 \\ 
2013-06-06 &  2456449.788 &  14.169 0.001 & $V$ &    1m0-4 & 2013-07-03 &  2456476.625 &  14.338 0.013 & $r$ &    1m0-4 \\ 
2013-06-07 &  2456450.793 &  14.160 0.019 & $V$ &    1m0-4 & 2013-07-07 &  2456480.625 &  14.460 0.005 & $r$ &    1m0-4 \\ 
2013-06-09 &  2456452.789 &  14.375 0.014 & $V$ &    1m0-4 & 2013-07-07 &  2456480.761 &  14.512 0.031 & $r$ &    1m0-5 \\ 
2013-06-12 &  2456455.789 &  14.333 0.003 & $V$ &    1m0-4 & 2013-07-09 &  2456482.754 &  14.566 0.022 & $r$ &    1m0-5 \\ 
2013-06-16 &  2456459.660 &  14.427 0.009 & $V$ &    1m0-9 & 2013-07-13 &  2456486.622 &  14.768 0.016 & $r$ &    1m0-4 \\ 
2013-06-20 &  2456463.661 &  14.449 0.010 & $V$ &    1m0-4 & 2013-07-14 &  2456487.622 &  15.031 0.060 & $r$ &    1m0-4 \\ 
2013-06-22 &  2456465.595 &  14.550 0.011 & $V$ &    1m0-4 & 2013-07-15 &  2456488.622 &  14.986 0.011 & $r$ &    1m0-4 \\ 
2013-06-24 &  2456467.595 &  14.523 0.003 & $V$ &    1m0-4 & 2013-07-31 &  2456504.695 &  16.317 0.011 & $r$ &    1m0-9 \\ 
2013-06-26 &  2456469.595 &  14.565 0.036 & $V$ &    1m0-4 & 2013-08-02 &  2456506.628 &  16.272 0.057 & $r$ &    1m0-5 \\ 
2013-06-28 &  2456471.595 &  14.642 0.008 & $V$ &    1m0-4 & 2013-08-09 &  2456513.605 &  16.359 0.007 & $r$ &    1m0-9 \\ 
2013-06-30 &  2456473.761 &  14.710 0.018 & $V$ &    1m0-5 & 2013-08-09 &  2456513.668 &  16.409 0.011 & $r$ &    1m0-4 \\ 
2013-07-02 &  2456475.619 &  14.848 0.006 & $V$ &    1m0-4 & 2013-08-10 &  2456515.498 &  16.488 0.013 & $r$ &    1m0-4 \\ 
2013-07-03 &  2456476.619 &  14.904 0.004 & $V$ &    1m0-4 & 2013-08-14 &  2456519.486 &  16.548 0.012 & $r$ &    1m0-9 \\ 
2013-07-07 &  2456480.619 &  15.092 0.007 & $V$ &    1m0-4 & 2013-08-16 &  2456520.601 &  16.561 0.034 & $r$ &    1m0-5 \\ 
2013-07-09 &  2456482.748 &  15.219 0.027 & $V$ &    1m0-5 & 2013-08-25 &  2456529.554 &  16.552 0.008 & $r$ &    1m0-9 \\ 
2013-07-13 &  2456486.616 &  15.550 0.014 & $V$ &    1m0-4 & 2013-08-28 &  2456532.530 &  16.631 0.001 & $r$ &    1m0-5 \\ 
2013-07-14 &  2456487.616 &  15.670 0.002 & $V$ &    1m0-4 & 2013-08-30 &  2456534.559 &  16.702 0.011 & $r$ &    1m0-4 \\ 
2013-07-15 &  2456488.616 &  15.772 0.006 & $V$ &    1m0-4 & 2013-09-21 &  2456556.521 &  16.910 0.008 & $r$ &    1m0-4 \\ 
2013-07-17 &  2456490.616 &  16.141 0.023 & $V$ &    1m0-4 & 2013-04-24 &  2456406.830 &  13.765 0.009 & $r$ &      CSP \\ 
2013-07-19 &  2456492.623 &  16.600 0.022 & $V$ &    1m0-4 & 2013-04-25 &  2456407.850 &  13.563 0.008 & $r$ &      CSP \\ 
2013-07-29 &  2456502.616 &  17.311 0.020 & $V$ &    1m0-4 & 2013-04-27 &  2456409.850 &  13.321 0.005 & $r$ &      CSP \\ 
2013-07-29 &  2456502.624 &  17.336 0.022 & $V$ &    1m0-9 & 2013-04-29 &  2456411.920 &  13.202 0.005 & $r$ &      CSP \\ 
2013-08-02 &  2456506.620 &  17.345 0.046 & $V$ &    1m0-5 & 2013-05-01 &  2456413.900 &  13.179 0.007 & $r$ &      CSP \\ 
2013-08-09 &  2456513.594 &  17.337 0.020 & $V$ &    1m0-9 & 2013-05-03 &  2456415.900 &  13.174 0.005 & $r$ &      CSP \\ 
2013-08-09 &  2456513.658 &  17.466 0.018 & $V$ &    1m0-4 & 2013-05-06 &  2456418.910 &  13.225 0.006 & $r$ &      CSP \\ 
2013-08-10 &  2456515.487 &  17.610 0.030 & $V$ &    1m0-4 & 2013-05-11 &  2456423.850 &  13.370 0.005 & $r$ &      CSP \\ 
2013-08-14 &  2456519.474 &  17.587 0.020 & $V$ &    1m0-9 & 2013-05-14 &  2456426.770 &  13.427 0.005 & $r$ &      CSP \\ 
2013-08-16 &  2456520.589 &  17.560 0.003 & $V$ &    1m0-5 & 2013-05-21 &  2456433.850 &  13.565 0.005 & $r$ &      CSP \\ 
2013-08-25 &  2456529.543 &  17.550 0.012 & $V$ &    1m0-9 & 2013-05-23 &  2456435.760 &  13.598 0.005 & $r$ &      CSP \\ 
2013-08-28 &  2456532.518 &  17.723 0.011 & $V$ &    1m0-5 & 2013-05-24 &  2456436.860 &  13.613 0.005 & $r$ &      CSP \\ 
2013-08-30 &  2456534.548 &  17.659 0.072 & $V$ &    1m0-4 & 2013-05-25 &  2456437.860 &  13.629 0.005 & $r$ &      CSP \\ 
2013-09-21 &  2456556.509 &  17.922 0.057 & $V$ &    1m0-4 & 2013-05-26 &  2456438.850 &  13.655 0.005 & $r$ &      CSP \\ 
2013-09-24 &  2456559.569 &  18.071 0.017 & $V$ &    1m0-4 & 2013-06-01 &  2456444.668 &  13.738 0.005 & $r$ &      CSP \\ 
2013-04-25 &  2456407.578 &  13.486 0.010 & $r$ &   1m0-10 & 2013-06-03 &  2456446.689 &  13.758 0.008 & $r$ &      CSP \\ 
2013-04-25 &  2456408.464 &  13.391 0.004 & $r$ &   1m0-10 & 2013-06-05 &  2456448.737 &  13.774 0.008 & $r$ &      CSP \\ 
2013-04-26 &  2456409.462 &  13.304 0.008 & $r$ &   1m0-10 & 2013-04-25 &  2456407.550 &  13.617 0.019 & $i$ &   1m0-10 \\ 
2013-04-27 &  2456410.463 &  13.314 0.014 & $r$ &   1m0-10 & 2013-04-25 &  2456408.465 &  13.429 0.002 & $i$ &   1m0-10 \\ 
2013-04-29 &  2456412.499 &  13.159 0.017 & $r$ &   1m0-10 & 2013-04-26 &  2456409.464 &  13.384 0.010 & $i$ &   1m0-10 \\ 
2013-05-03 &  2456415.785 &  13.120 0.007 & $r$ &    1m0-4 & 2013-04-27 &  2456410.464 &  13.300 0.021 & $i$ &   1m0-10 \\ 
2013-05-03 &  2456415.864 &  13.177 0.002 & $r$ &    1m0-5 & 2013-04-28 &  2456411.464 &  13.165 0.008 & $i$ &   1m0-10 \\ 
2013-05-04 &  2456416.823 &  13.167 0.034 & $r$ &    1m0-5 & 2013-04-30 &  2456412.512 &  13.174 0.009 & $i$ &   1m0-10 \\ 
2013-05-05 &  2456417.786 &  13.121 0.011 & $r$ &    1m0-4 & 2013-05-03 &  2456415.786 &  13.121 0.071 & $i$ &    1m0-4 \\ 
2013-05-06 &  2456418.785 &  13.180 0.026 & $r$ &    1m0-4 & 2013-05-03 &  2456415.869 &  13.131 0.006 & $i$ &    1m0-5 \\ 
2013-05-11 &  2456423.749 &  13.355 0.004 & $r$ &    1m0-9 & 2013-05-04 &  2456416.828 &  13.135 0.030 & $i$ &    1m0-5 \\ 
2013-05-15 &  2456427.748 &  13.482 0.005 & $r$ &    1m0-9 & 2013-05-05 &  2456417.787 &  13.171 0.008 & $i$ &    1m0-4 \\ 
2013-05-31 &  2456443.720 &  13.678 0.009 & $r$ &    1m0-9 & 2013-05-06 &  2456418.786 &  13.111 0.016 & $i$ &    1m0-4 \\ 
2013-06-03 &  2456446.793 &  13.739 0.009 & $r$ &    1m0-4 & 2013-05-31 &  2456443.723 &  13.650 0.007 & $i$ &    1m0-9 \\ 
2013-06-04 &  2456447.793 &  13.726 0.001 & $r$ &    1m0-4 & 2013-06-03 &  2456446.795 &  13.700 0.008 & $i$ &    1m0-4 \\ 
2013-06-05 &  2456448.793 &  13.792 0.009 & $r$ &    1m0-4 & 2013-06-04 &  2456447.795 &  13.758 0.019 & $i$ &    1m0-4 \\ 
2013-06-07 &  2456450.798 &  13.764 0.012 & $r$ &    1m0-4 & 2013-06-05 &  2456448.796 &  13.683 0.023 & $i$ &    1m0-4 \\ 
2013-06-09 &  2456452.795 &  13.777 0.012 & $r$ &    1m0-4 & 2013-06-09 &  2456452.797 &  13.851 0.013 & $i$ &    1m0-4 \\ 
2013-06-12 &  2456455.795 &  13.849 0.002 & $r$ &    1m0-4 & 2013-06-12 &  2456455.797 &  13.818 0.010 & $i$ &    1m0-4 \\ 
2013-06-16 &  2456459.671 &  13.913 0.010 & $r$ &    1m0-9 & 2013-06-22 &  2456465.603 &  13.951 0.013 & $i$ &    1m0-4 \\ 
2013-06-20 &  2456463.671 &  14.005 0.006 & $r$ &    1m0-4 & 2013-06-24 &  2456467.603 &  14.005 0.005 & $i$ &    1m0-4 \\ 
2013-06-22 &  2456465.601 &  14.021 0.018 & $r$ &    1m0-4 & 2013-06-26 &  2456469.603 &  14.050 0.002 & $i$ &    1m0-4 \\ 
2013-06-24 &  2456467.601 &  14.076 0.016 & $r$ &    1m0-4 & 2013-06-28 &  2456471.603 &  14.073 0.005 & $i$ &    1m0-4 \\ 
2013-06-26 &  2456469.601 &  14.063 0.010 & $r$ &    1m0-4 & 2013-06-30 &  2456473.768 &  14.152 0.017 & $i$ &    1m0-5 \\ 
\hline
    \end{tabular}
    \label{tab2}
    \end{minipage}
    \end{table*}
    
\begin{table*}
    \centering
    \begin{minipage}{140mm}
    \caption{Photometric Data}
    \begin{tabular}{ccccc|ccccc}
    \hline 
Date &JD &mag &Filter &telescope &Date &JD &mag &Filter &telescope \footnote{{\em Swift} Telescope; CSP (Las Campanas, Chile); 1m0-08 (McDonald observatory, USA); 1m0-10, 1m0-12, 1m0-13 (Sutherland, South Africa), 1m0-04, 1m0-05, 1m0-09  (Cerro Tololo, Chile); 1m0-03, 1m0-11 (Siding Spring, Australia)}\\ 
 \hline 
2013-07-02 &  2456475.628 &  14.220 0.021 & $i$ &    1m0-4 & 2013-04-24 &  2456406.830 &  13.916 0.013 & $i$ &      CSP \\ 
2013-07-03 &  2456476.628 &  14.290 0.008 & $i$ &    1m0-4 & 2013-04-25 &  2456407.850 &  13.663 0.008 & $i$ &      CSP \\ 
2013-07-07 &  2456480.627 &  14.413 0.003 & $i$ &    1m0-4 & 2013-04-27 &  2456409.840 &  13.371 0.006 & $i$ &      CSP \\ 
2013-07-09 &  2456482.756 &  14.538 0.007 & $i$ &    1m0-5 & 2013-04-29 &  2456411.920 &  13.216 0.005 & $i$ &      CSP \\ 
2013-07-13 &  2456486.624 &  14.809 0.013 & $i$ &    1m0-4 & 2013-05-01 &  2456413.900 &  13.145 0.005 & $i$ &      CSP \\ 
2013-07-14 &  2456487.624 &  14.900 0.008 & $i$ &    1m0-4 & 2013-05-03 &  2456415.900 &  13.127 0.005 & $i$ &      CSP \\ 
2013-07-15 &  2456488.624 &  15.091 0.009 & $i$ &    1m0-4 & 2013-05-06 &  2456418.910 &  13.181 0.006 & $i$ &      CSP \\ 
2013-08-02 &  2456506.631 &  16.376 0.006 & $i$ &    1m0-5 & 2013-05-11 &  2456423.850 &  13.351 0.006 & $i$ &      CSP \\ 
2013-08-09 &  2456513.608 &  16.580 0.014 & $i$ &    1m0-9 & 2013-05-14 &  2456426.770 &  13.420 0.005 & $i$ &      CSP \\ 
2013-08-09 &  2456513.674 &  16.544 0.029 & $i$ &    1m0-4 & 2013-05-21 &  2456433.850 &  13.570 0.005 & $i$ &      CSP \\ 
2013-08-11 &  2456515.503 &  16.601 0.041 & $i$ &    1m0-4 & 2013-05-23 &  2456435.770 &  13.599 0.005 & $i$ &      CSP \\ 
2013-08-14 &  2456519.491 &  16.711 0.006 & $i$ &    1m0-9 & 2013-05-24 &  2456436.860 &  13.610 0.006 & $i$ &      CSP \\ 
2013-08-16 &  2456520.605 &  16.677 0.020 & $i$ &    1m0-5 & 2013-05-25 &  2456437.860 &  13.632 0.005 & $i$ &      CSP \\ 
2013-08-25 &  2456529.558 &  16.723 0.004 & $i$ &    1m0-9 & 2013-05-26 &  2456438.860 &  13.657 0.005 & $i$ &      CSP \\ 
2013-08-28 &  2456532.534 &  16.835 0.007 & $i$ &    1m0-5 & 2013-06-01 &  2456444.669 &  13.754 0.005 & $i$ &      CSP \\ 
2013-08-30 &  2456534.564 &  16.927 0.001 & $i$ &    1m0-4 & 2013-06-03 &  2456446.691 &  13.754 0.008 & $i$ &      CSP \\ 
2013-09-21 &  2456556.525 &  17.170 0.011 & $i$ &    1m0-4 & 2013-06-05 &  2456448.739 &  13.787 0.006 & $i$ &      CSP \\ 
\hline
    \end{tabular}
    \label{tab2}
    \end{minipage}
    \end{table*}

\end{document}